\documentclass[12pt]{article}
\newcommand{\be}{\begin{eqnarray}}\newcommand{\ee}{\end{eqnarray}}
\newcommand{\beo}{\begin{eqnarray*}}\newcommand{\eeo}{\end{eqnarray*}}
\newcommand{\ba}{\begin{array}}\newcommand{\ea}{\end{array}}
\newcommand{\no}{\nonumber}

\begin{document}

\title{\textbf{Background Independent Open String\\Field Theory with Constant $B$ Field\\On the Annulus}}
\author{R. Rashkov\footnote{rash@phys.uni-sofia.bg}, K. S. Viswanathan\footnote{kviswana@sfu.ca} and Y. Yang\footnote{yyangc@sfu.ca}}
\date{\textit{Department of Physics\\Simon Fraser University\\Vancouver, BC V5A 1S6 Canada}}
\maketitle

\begin{abstract}
We study Witten's background independent open-string field theory in the presence of a constant
$B$-field at one loop level. The Green's function and the partition function with a constant
$B$-field are evaluated for an annulus.
\end{abstract}

\section{Introduction}

Background independent open-string field theory has been formally defined in the space of all
two-dimensional world-sheet theories by Witten \cite{witten1}. The first concrete computation
of the action for an off-shell tachyon field in the tree level was given in \cite{witten2}.
It has been shown recently that by turning on a large $B$-field, the noncommutative geometry
\cite{witten3} simplifies the construction of soliton solution \cite{strominger} which leads
to tachyon condensation \cite{rajesh}. The stationery point of the tachyon potential on
the disk from background independent open string field theory \cite{witten2,shatashvili1}
agrees precisely with the tension of the D-brane \cite{shatashvili2,moore,sen}. The tachyon
potential with constant $B$-field on the disk was constructed in \cite{yasnov}. The disc
partition function for the D0-D2 system with a large $B$-field was evaluated in \cite{david}.
Since at the tree level the tachyon field seems to be playing the role of the loop counting
parameter, it would be interesting to see the form of the exponential in the tachyon potential
in higher order terms. In this letter we generalize the calculations of the partition function
in \cite{witten2,yasnov} beyond the tree level by including the annulus diagram.\\

The organization for this paper is follows. In section 2 we define the boundary value problem
of background independent open-string field theory with a $B$-field on the annulus. In section
3 we obtain the Green's function on the annulus. In section 4 we use the Green's function to
evaluate the partition function on the annulus in presence of $B$-field. Section 5 contain a 
brief discusion.\\

\section{The Boundary Problem on the Annulus}

Following \cite{witten2,yasnov} we consider a two-dimensional action of the form

\be
I=I_0+I'
\ee
where $I_0$ is the bulk action and the boundary term $I'$ describes coupling of the external
open strings. The bulk action $I_0$ describes the closed string background with a constant
$B$-field

\be
I_0=\int_\Sigma d^2x\sqrt{h}\left(\frac{1}{8\pi}h^{ij}\partial_iX^\mu\partial_jX_\mu
+b^{ij}D_ic_j\right)
-\frac{i}{8\pi}\int_\Sigma d^2x\epsilon^{ij}B_{\mu\nu}\partial_iX^\mu\partial_jX^\nu
\ee
where $c_I$ and $b^{ij}$ are the ghost and anti-ghost fields. $\Sigma$ is any two-dimensional
world sheet with the metric $h_{ij}$ and coordinates $X^i$. In this letter we choose $\Sigma$
to be the annulus with a rotationally invariant flat metric

\be
ds^2=d\sigma_1^2+d\sigma_2^2 \hspace{1.2cm} a\le\sigma_1^2+\sigma_2^2\le b
\ee
It is convenient to work with complex coordinates and set $z=\sigma_1+i\sigma_2$, with
$a\le|z|\le b$\\

The boundary term $I'$ can be any ghost number conserving boundary interaction, for which
everything can be computed explicitly, we will take it to be a quadratic function of the
coordinates

\be
I'=\frac{1}{8\pi}\int_{\partial\Sigma}d\theta u_\mu (X^\mu)^2
\ee
The boundary condition derived by varying $I=I_0+I'$ is

\be
(1+B)_{\mu\nu}z\partial X^\nu+(1-B)_{\mu\nu}\bar z\bar\partial X^\nu
+u_\mu X^\mu|_{\partial\Sigma}=0
\ee
Note there is no sum on $\mu$ in the last term.\\

The Green's function of the theory should obey

\be
\partial_z \bar{\partial}_{\bar{z}} G(z,w) = -2 \pi\alpha' \delta^{(2)} (z-w)
\ee
with the boundary condition (5).\\

The Green's function for the case without boundary term ($u=0$) has been solved in
\cite{callan}. In that case the boundary conditions are

\be
\ba{l} (1+B)_{\mu\nu}z\partial G^\nu_\lambda+(1-B)_{\mu\nu}\bar z\bar\partial G^\nu_\lambda
|_{r=b}=-\alpha'\delta_{\mu\lambda}\\
(1+B)_{\mu\nu}z\partial G^\nu_\lambda+(1-B)_{\mu\nu}\bar z\bar\partial G^\nu_\lambda|_{r=a}=0
\ea
\ee
The boundary conditions can not be set to 0 for both $r=a$ and $r=b$, because in the presence
of the $B$-field, eq. (6) and Gauss's theorem would not permit the propagator to be
single-valued. The simplest choice is (7).\\

In our case with the $u\not =0$ term, we have the freedom to set both boundary conditions to
0, i.e.

\be
\ba{l} (1+B)_{\mu\nu}z\partial X^\nu+(1-B)_{\mu\nu}\bar z\bar\partial X^\nu+u_\mu X^\mu|_{r=b}=0
\\
(1+B)_{\mu\nu}z\partial X^\nu+(1-B)_{\mu\nu}\bar z\bar\partial X^\nu+u_\mu X^\mu|_{r=a}=0
\ea
\ee
provided 

\be
\int_{r=a,b}\frac{uG}{r}ds=2\pi\alpha'
\ee

\section{The Green's Function}

To solve the eq. (6) with the boundary conditons (8), we start with the ansatz,

\be
G(z,w)& = &\frac{\alpha'}{2}\left\{-ln|z-w|^2
+\frac{[ln(z\bar z/b^2)-2/u][ln (w\bar w/b^2)-2/u]}{ln(a^2/b^2)}\right.\no\\
& &+\sum_{k=-\infty}^{\infty}[a_k (z\bar w)^k+a'_k(\bar z w)^k]
+\left.\sum_{k=-\infty}^{\infty}\left[b_k (z/w)^k+ b'_k(\bar z/\bar w)^k\right]\right\}
\ee
The second term is added because $ln(z\bar z)$ solves eq. (6) on annulus. The pre-factor
$\alpha'/2$ makes the Green's function satisfy the eq. (9). $a_k$'s and $b_k$'s are the
coefficents to be determinded by the boundary conditions (8).\\

Inserting this ansatz into the boundary conditions (8) and expanding it by series, we get

\be
0&=&(z \partial + \bar{z} \bar{\partial}+u)G|_{r=a}\no\\
&=&\sum_{k=-\infty}^{-1}\left(\frac{1+B}{a^{2k}}-\frac{u}{ka^{2k}}+[(1-B)k+u]a'_k
-\frac{[(1+B)k-u]b_{-k}}{a^{2k}}\right)(\bar z w)^k\no\\
& &+\sum_{k=-\infty}^{-1}\left(\frac{1-B}{a^{2k}}-\frac{u}{ka^{2k}}+[(1+B)k+u]a_k
-\frac{[(1-B)k-u]b'_{-k}}{a^{2k}}\right)(z\bar w)^k\no\\
& &+\sum_{k=1}^{\infty}\left([(1-B)k+u]a'_k-\frac{[(1+B)k-u]b_{-k}}{a^{2k}}\right)(\bar z w)^k\\
& &+\sum_{k=1}^{\infty}\left([(1+B)k+u]a_k-\frac{[(1-B)k-u]b'_{-k}}{a^{2k}}\right)(z\bar w)^k\no\\
& &+u(a_0+a'_0+b_0+b'_0)-2-ulnb^2\no\\
\no\\
0&=&(z \partial + \bar{z} \bar{\partial}+u)G|_{r=a}\no\\
&=&\sum_{k=1}^{\infty}\left(-\frac{1+B}{b^{2k}}+\frac{u}{kb^{2k}}+[(1-B)k+u]a'_k
-\frac{[(1+B)k-u]b_{-k}}{b^{2k}}\right)(\bar z w)^k\no\\
& &+\sum_{k=1}^{\infty}\left(-\frac{1-B}{b^{2k}}+\frac{u}{kb^{2k}}+[(1+B)k+u]a_k
-\frac{[(1-B)k-u]b'_{-k}}{b^{2k}}\right)(z\bar w)^k\no\\
& &+\sum_{k=-\infty}^{-1}\left([(1-B)k+u]a'_k-\frac{[(1+B)k-u]b_{-k}}{b^{2k}}\right)(\bar z w)^k\no\\
& &+\sum_{k=-\infty}^{-1}\left([(1+B)k+u]a_k-\frac{[(1-B)k-u]b'_{-k}}{b^{2k}}\right)(z\bar w)^k\\
& &+u(a_0+a'_0+b_0+b'_0)-2-ulnb^2\no
\ee
namely,

\be
u(a_0+a'_0+b_0+b'_0)-2-ulnb^2=0
\ee

\be
\left\{
\ba{r}\frac{1+B}{a^{2k}}-\frac{u}{ka^{2k}}+[(1-B)k+u]a'_k-\frac{[(1+B)k-u]b_{-k}}{a^{2k}}=0\\
\frac{1-B}{a^{2k}}-\frac{u}{ka^{2k}}+[(1+B)k+u]a_k-\frac{[(1-B)k-u]b'_{-k}}{a^{2k}}=0\\
\left[(1-B)k+u\right]a'_k-\frac{[(1+B)k-u]b_{-k}}{b^{2k}}=0\\
\left[(1+B)k+u\right]a_k-\frac{[(1-B)k-u]b'_{-k}}{b^{2k}}=0
\ea\mbox{ for $k<0$}\right.
\ee

\be
\left\{
\ba{r}-\frac{1+B}{b^{2k}}+\frac{u}{kb^{2k}}+[(1-B)k+u]a'_k-\frac{[(1+B)k-u]b_{-k}}{b^{2k}}=0\\
-\frac{1-B}{b^{2k}}+\frac{u}{kb^{2k}}+[(1+B)k+u]a_k-\frac{[(1-B)k-u]b'_{-k}}{b^{2k}}=0\\
\left[(1-B)k+u\right]a'_k-\frac{[(1+B)k-u]b_{-k}}{a^{2k}}=0\\
\left[(1+B)k+u\right]a_k-\frac{[(1-B)k-u]b'_{-k}}{a^{2k}}=0
\ea\mbox{ for $k>0$}\right.
\ee
the solutions are

\be
a_0+a'_0+b_0+b'_0=\frac{2}{u}+lnb^2
\ee

\be
\left\{
\ba{l}a_k=\frac{1-B}{1+B}\frac{1}{k(b^{2k}-a^{2k})}
-\frac{2u}{k(1+B)}\frac{1}{(1+B)k+u}\frac{1}{k(b^{2k}-a^{2k})}\\
a'_k=\frac{1+B}{1-B}\frac{1}{k(b^{2k}-a^{2k})}
-\frac{2u}{k(1-B)}\frac{1}{(1-B)k+u}\frac{1}{k(b^{2k}-a^{2k})}\\
b_{-k}=\frac{b^{2k}}{k(b^{2k}-a^{2k})}
\Rightarrow b_k=\frac{a^{2k}}{k(b^{2k}-a^{2k})}\\
b'_{-k}=\frac{b^{2k}}{k(b^{2k}-a^{2k})}
\Rightarrow b'_k=\frac{a^{2k}}{k(b^{2k}-a^{2k})}\ea
\mbox{ for $k<0$}\right.
\ee

\be
\left\{
\ba{l}a_k=\frac{1-B}{1+B}\frac{1}{k(b^{2k}-a^{2k})}
-\frac{2u}{k(1+B)}\frac{1}{(1+B)k+u}\frac{1}{k(b^{2k}-a^{2k})}\\
a'_k=\frac{1+B}{1-B}\frac{1}{k(b^{2k}-a^{2k})}
-\frac{2u}{k(1-B)}\frac{1}{(1-B)k+u}\frac{1}{k(b^{2k}-a^{2k})}\\
b_{-k}=\frac{a^{2k}}{k(b^{2k}-a^{2k})}
\Rightarrow b_k=\frac{b^{2k}}{k(b^{2k}-a^{2k})}\\
b'_{-k}=\frac{a^{2k}}{k(b^{2k}-a^{2k})}
\Rightarrow b'_k=\frac{b^{2k}}{k(b^{2k}-a^{2k})}\ea
\mbox{ for $k>0$}\right.
\ee

Inserting the coefficients $a_k$, $a'_k$, $b_k$ and $b'_k$ back into the ansatz, we obtain
the Green's function on the annulus, which is 

\be
G(z,w)&=&\frac{\alpha'}{2}\left \{-ln|z-w|^2+\frac{2}{u}+lnb^2\right.\no\\
& &+\frac{[ln(z\bar z/b^2)-2/u][ln (w\bar w/b^2)-2/u]}{ln(a^2/b^2)}\no\\
& &-\sum_{n=1}^{\infty}\left[ln\left |1-\left(\frac{a}{b}\right)^{2n}\frac{z}{w}\right |^2
   +ln\left |1-\left(\frac{a}{b}\right)^{2n}\frac{w}{z}\right |^2\right ]\\
& &-\frac{1-B}{1+B}\sum_{n=1}^{\infty}\left [ln\left(1-\left(\frac{a}{b}\right)^{2n}\frac{b^2}{z\bar{w}}\right )+ln\left (1-\left(\frac{a}{b}\right)^{2n} \frac{z\bar{w}}{a^2}\right )\right]\no\\
& &-\frac{1+B}{1-B}\sum_{n=1}^{\infty}\left [ln\left(1-\left(\frac{a}{b}\right)^{2n}\frac{b^2}{\bar z w}\right )+ln\left (1-\left(\frac{a}{b}\right)^{2n} \frac{\bar z w}{a^2}\right )\right]\no\\
& &+\sum_{k=1}^{\infty}\frac{2u}{k(1+B)}\frac{1}{(1+B)k-u}\frac{a^{2k}}{b^{2k}-a^{2k}}\left(\frac{b^2}{z\bar{w}}\right)^k\no\\
& &+\sum_{k=1}^{\infty}\frac{2u}{k(1-B)}\frac{1}{(1-B)k-u}\frac{a^{2k}}{b^{2k}-a^{2k}}\left(\frac{b^2}{\bar z w}\right)^k\no\\
& &-\sum_{k=1}^{\infty}\frac{2u}{k(1+B)}\frac{1}{(1+B)k+u}\frac{a^{2k}}{b^{2k}-a^{2k}}\left(\frac{z\bar{w}}{a^2}\right)^k\no\\
& &-\sum_{k=1}^{\infty}\frac{2u}{k(1-B)}\frac{1}{(1-B)k+u}\frac{a^{2k}}{b^{2k}-a^{2k}}\left(\frac{\bar z w}{a^2}\right)^k
\ee
Note that $u=0$ is the singular point for this Green's function, which matches the result in
\cite{witten2}. But instead of $\sim 1/u$  behavior on the disc, the Green's function on the
annulus behaves $\sim 1/u^2$. It's natural to conjecture that for the world sheet with $N$
boundaries, the Green's function behaves $\sim 1/u^N$.

\section{The Partition Function}
We will use the Green's function which we found above to evaluate the partition function on
the annulus as follows. Define

\be
\langle X(\theta)X(\theta)\rangle |_{\partial\Sigma}=
\lim_{\epsilon\to 0}[X(\theta)X(\theta)-f(\epsilon)]
\ee
where

\be
f(\epsilon)=-\frac{2}{1+B}ln(1-e^{i\epsilon})-\frac{2}{1-B}ln(1-e^{-i\epsilon})
\ee
To simplify our presentation we consider the case where the only non-zero components of the
$B$-field are $B_{12}=-B_{21}=b_B$. The result can be generalized easily as below.
In this case we have two scalar fields $X^1$ and $X^2$ with $u_1=u_2=u$. With the above
definitions, we have

\be
\frac{2}{\alpha'}\langle X_1(\theta)X_1(\theta)\rangle|_{r=b}
&=&\frac{2}{\alpha'}\langle X_2(\theta)X_2(\theta)\rangle|_{r=b}\no\\
&=&-\frac{4}{1+b_B^2}\sum_{n=1}^{\infty}ln\left|1-\left(\frac{a}{b}\right)^{2n}\right|^2
+\frac{2}{u}+\frac{4}{u^2 ln(a^2/b^2)}\no\\
& &+\frac{4u}{1+b_B^2}\sum_{k=1}^{\infty}\frac{1}{k}
\frac{k-u-b_B^2k}{(k-u)^2+b_B^2k^2}\frac{a^{2k}}{b^{2k}-a^{2k}}\\
& &-\frac{4u}{1+b_B^2}\sum_{k=1}^{\infty}\frac{1}{k}
\frac{k+u-b_B^2k}{(k+u)^2+b_B^2k^2}\frac{b^{2k}}{b^{2k}-a^{2k}}\no\\
\no\\\no\\
\frac{2}{\alpha'}\langle X_1(\theta)X_1(\theta)\rangle|_{r=a}
&=&\frac{2}{\alpha'}\langle X_2(\theta)X_2(\theta)\rangle|_{r=a}\no\\
&=&-\frac{4}{1+b_B^2}\sum_{n=1}^{\infty}ln\left|1-\left(\frac{a}{b}\right)^{2n}\right|^2
-\frac{2}{u}+\frac{4}{u^2 ln(a^2/b^2)}\no\\
& &+\frac{4u}{1+b_B^2}\sum_{k=1}^{\infty}\frac{1}{k}
\frac{k-u-b_B^2k}{(k-u)^2+b_B^2k^2}\frac{b^{2k}}{b^{2k}-a^{2k}}\\
& &-\frac{4u}{1+b_B^2}\sum_{k=1}^{\infty}\frac{1}{k}
\frac{k+u-b_B^2k}{(k+u)^2+b_B^2k^2}\frac{a^{2k}}{b^{2k}-a^{2k}}\no
\ee

\be
&\Rightarrow&\frac{d}{du}lnZ(a/b)\no\\
&=&-\frac{1}{8\pi}\left [\int_0^{2\pi}\langle X(\theta)X(\theta)\rangle |_{r=b}d\theta
-\int_0^{2\pi}\langle X(\theta)X(\theta)\rangle |_{r=a}d\theta\right ]\no\\
&=&-\frac{\alpha'}{4}\left [-\frac{4u}{1+b_B^2}\sum_{k=1}^{\infty}\frac{1}{k}\frac{k-u-b_B^2k}{(k-u)^2+b_B^2k^2}
-\frac{4u}{1+b_B^2}\sum_{k=1}^{\infty}\frac{1}{k}\frac{k+u-b_B^2k}{(k+u)^2+b_B^2k^2}\right]+\frac{4}{u}\no\\
&=&\frac{\alpha'}{2}\left[\frac{1}{1+ib_B}\psi(1+\frac{u}{1+ib_B})+\frac{1}{1-ib_B}\psi(1+\frac{u}{1-ib_B})\right.\no\\
& &\left.-\frac{1}{1-ib_B}\psi(1-\frac{u}{1-ib_B})-\frac{1}{1+ib_B}\psi(1-\frac{u}{1+ib_B})-\frac{2}{u}\right]\no\\
&\equiv&\frac{\alpha'}{2}\left [\frac{d}{du}ln\Gamma(1+\frac{u}{1+ib_B})+\frac{d}{du}ln\Gamma(1+\frac{u}{1-ib_B})\right.\no\\
& &+\left.\frac{d}{du}ln\Gamma(1-\frac{u}{1-ib_B})+\frac{d}{du}ln\Gamma(1-\frac{u}{1+ib_B})-\frac{2}{u}\right]
\ee

\be
\Rightarrow Z(a/b)=\left[\frac{\Gamma(1+\frac{u}{1+ib_B})\Gamma(1+\frac{u}{1-ib_B})\Gamma(1-\frac{u}{1+ib_B})\Gamma(1-\frac{u}{1-ib_B})}{u^2}\right]^{\alpha'/2}
\ee
It is a surpise that the partition function on the annulus does not depend on the
modulus of the annulus. We believe that there must be a deep reason which we do not understand
right now. To include all topologically different annuli, we need to integrate over the ratio
$a/b$ from 0 to 1. It can be trivially done as

\be
\Rightarrow Z=\int_0^1 Z(a/b)\,d(a/b)=\left[\frac{\Gamma(1+\frac{u}{1+ib_B})
\Gamma(1+\frac{u}{1-ib_B})\Gamma(1-\frac{u}{1+ib_B})\Gamma(1-\frac{u}{1-ib_B})}{u^2}\right]^{\alpha'/2}
\ee
Having obtained the partition function for the special case of a single non-vanishing
component of the $B$-field, it is easy to determine the partition function for the general case.
Using the formula

\be
det\Gamma\left(1+\frac{u}{1+B}\right)=det\Gamma\left(1+\frac{u}{1-B}\right)
=\Gamma\left(1+\frac{u}{1+ib_B}\right)\Gamma\left(1+\frac{u}{1-ib_B}\right)\no\\
& &\\
det\Gamma\left(1-\frac{u}{1+B}\right)=det\Gamma\left(1-\frac{u}{1-B}\right)
=\Gamma\left(1-\frac{u}{1+ib_B}\right)\Gamma\left(1-\frac{u}{1-ib_B}\right)\no
\ee
We obtain our final result

\be
Z=\left\{\frac{det\left[\Gamma(1+\frac{u}{1+B})
\Gamma(1+\frac{u}{1-B})\Gamma(1-\frac{u}{1+B})
\Gamma(1-\frac{u}{1-B})\right]}{u^4}\right\}^{\alpha'/4}
\ee
When $\alpha'=1$, $Z\sim 1/u=1/(\sqrt{u})^2$ around $u=0$, which agrees with the analysis
in \cite{witten2}.

\section{Discussion}
We have obtained the partition function with $B$-field on the annulus. Compare to the partition
function on the disk, there are some new features which need to further understand.

\begin{itemize}
\item
The Green's function on the annulus has a term $\sim 1/u^2$ instead of $\sim 1/u$ on
the disc. So its natural to ask whether the Green's function on a world sheet with $N$
boundaries will have a term $\sim 1/u^N$.
\item
The partition function on the annulus is modulus independent, there must be a deep
reason for it.
\item
In eq. (27), if $b_B\not=0$, there is no singularity for any real value of $u$.
But if $b_B=0$, then all of the integer points turn to be singular points. The question is what
physical meaning of those singular points.
\end{itemize}

Using one loop effective tachyon action to understand tachyon condensation is currently under
investigation.

\end{document}